\documentclass{PoS}

\title{Direct reactions in/for astrophysics}

\ShortTitle{Direct reactions in/for astrophysics}

\author{\speaker{Carlos Bertulani}%
         \thanks{I thank A. Mengoni and C. Spitaleri for beneficial
discussions. This work was
supported by the U.S. Department of Energy under Grant
No. DE-FG02-04ER41338.}\\
        Department of Physics, University of Arizona,  Tucson, AZ 85721\\
        E-mail: \email{bertulani@physics.arizona.edu}}

\abstract{Precise nuclear reaction rates are needed for a detailed
description of the production of elements in primordial
nucleosynthesis and during the hydrostatic burning of stars to
constrain the astrophysical models. The relevant reactions are
extremely difficult to measure directly in the laboratory at the
small astrophysical energies. In recent years direct reactions
methods have been developed and applied to extract low-energy
astrophysical S-factors. The application of these methods requires a
combination of experimental and theoretical efforts. This
contribution focuses on the underlying reaction theories that have
to be well understood in order to assess the precision and
limitations of the various approaches.}

\FullConference{International Symposium on Nuclear Astrophysics - Nuclei in the Cosmos - IX\\
         25-30 June 2006\\
         CERN}

\begin{document}

\section{How difficult is nuclear astrophysics?}
Ongoing studies in nuclear astrophysics are focused on the opposite ends of
the energy scale of nuclear reactions: (a) the very high and (b) the very low
relative energies between the reacting nuclei. Projectiles with high
bombarding energies produce nuclear matter at high densities and temperatures.
One expects that matter produced in central nuclear collisions will undergo a
phase transition and produce a quark-gluon
plasma. One can thus reproduce conditions existing in the first seconds of
the universe and also in the core of neutron stars. At the other end of the
energy scale are the low energy reactions of importance for stellar evolution.
Chains of nuclear reactions
lead to complicated phenomena like nucleosynthesis, supernovae explosions, and
energy production in stars.

In our Sun the reaction $^{7}$Be$\left(  {\rm p},\gamma\right)  ^{8}$B plays
a major role for the production of high energy neutrinos
from the $\beta$-decay of $^{8}$B. These neutrinos come directly from center
of the Sun and are ideal probes of the sun's structure. {\it John Bahcall} frequently
said that this was the most important reaction in nuclear astrophysics.
Our knowledge about this reaction has improved considerably due to the appearance of
radioactive beam facilities.
The reaction $^{12}$C$\left(  \alpha,\gamma\right)  ^{16}$O is extremely
relevant for the fate of massive stars. It determines if the remnant of a
supernova explosion becomes a black-hole or a neutron star. These two reactions are
only two examples of a large number of reactions which are not yet known with
the required accuracy needed in
astrophysics.

Approximately half of all stable nuclei observed in nature in the heavy
element region, $A>60$, are produced in the r--process. This r--process
occurs in environments with large neutron densities which leads to neutron
capture times much smaller than the beta-decay half--lives, $\;\tau
_{\mathrm{n}}\ll\tau_{\beta}$. The most neutron--rich isotopes along the
r--process path have lifetimes of less than one second; typically 10$^{-2}$ to
10$^{-1}$\thinspace s. Cross sections for most of the nuclei involved are hard
to measure experimentally. Sometimes, theoretical calculations of the capture
cross sections as well as the beta--decay half--lives are the only source of
the nuclear physics input for r--process calculations.

Nucleosynthesis in stars is complicated by the
presence of electrons. They screen the nuclear charges, therefore increasing
the fusion probability by reducing the Coulomb repulsion. Evidently, the
fusion cross sections measured in the laboratory have to be corrected by the
electron screening when used in a stellar model. This is a purely
theoretical problem as one can not reproduce the interior of stars in the
laboratory.

A simpler screening mechanism occurs in laboratory experiments due to the
bound atomic electrons in the nuclear targets. This case has been studied in
great details experimentally, as one can control different charge states of
the projectile+target system in the laboratory
\cite{Ass87,Rol95,Rol01}. The experimental findings disagree
systematically by a factor of two with theory. This is surprising as the
theory for atomic screening in the laboratory relies on our basic knowledge of
atomic physics. At very low energies one can use the simple adiabatic model in
which the atomic electrons rapidly adjust their orbits to the relative motion
between the nuclei prior to the fusion process. Energy conservation requires
that the larger electronic binding (due to a larger charge of the combined
system) leads to an increase of the relative motion between the nuclei, thus
increasing the fusion cross section. As a matter of fact, this enhancement has
been observed experimentally. The measured values are however not compatible
with the adiabatic estimate \cite{Ass87,Rol95,Rol01}. Dynamical
calculations have been performed, but they obviously cannot explain the
discrepancy as they include atomic excitations and ionizations which reduce
the energy available for fusion. Other small effects, like vacuum
polarization, atomic and nuclear polarizabilities, relativistic effects, etc.,
have also been considered \cite{BBH97}. But the discrepancy between experiment
and theory remains \cite{BBH97,Rol01}.

A possible solution of the laboratory screening problem was proposed in refs.
\cite{LSBR96,BFMH96}. Experimentalists often use the extrapolation of the
{\it Andersen-Ziegler} tables \cite{AZ77} to obtain the average value of the
projectile energy due to stopping in the target material. The stopping is due
to ionization, electron-exchange, and other atomic mechanisms. However, the
extrapolation is challenged by theoretical calculations which predict a lower
stopping. Smaller stopping was indeed verified experimentally \cite{Rol01}. At
very low energies, it is thought that the stopping mechanism is mainly due to
electron exchange between projectile and target. This has been studied in ref.
\cite{BD00} in the simplest situation; proton+hydrogen collisions.
The calculated stopping power was added
to the nuclear stopping power mechanism, i.e. to the energy loss by the
Coulomb repulsion between the nuclei. The obtained stopping power is
proportional to $v^{\alpha}$, where $v$ is the projectile velocity and
$\alpha=1.35$. The extrapolations from the Andersen-Ziegler table predict a
smaller value of $\alpha$. Although this result seems to indicate the stopping
mechanism as a possible reason for the laboratory screening problem, the
theoretical calculations tend to disagree on the power of $v$ at low energy
collisions. For example, ref. \cite{GS91} found $S\sim v_{p}^{3.34}$ for
protons in the energy range of 4 keV incident on helium targets. This is an
even larger deviation from the extrapolations of the Andersen-Ziegler tables.

Another calculation of the stopping power in
atomic He$^{+}+$He collisions using the two-center molecular orbital basis
was done in ref. \cite{Ber04}.  The
agreement with the data from ref. \cite{GS91} at low energies is excellent.
The
agreement with the data disappears completely if the nuclear recoil is
included. In fact, the unexpected "disappearance" of the nuclear recoil was
also observed in ref. \cite{Form}. This seems to violate a
basic principle of nature, as the nuclear recoil is due to the Coulomb
repulsion between the projectile and the target atoms \cite{AZ77}. Much of
what we know in this field now is due to the work of {\it Claus Rolfs,
Karlheinz Langanke, Noboru Takigawa, Kouichi Hagino, Baha Balantekin}, and
collaborators.

In the previous paragraphs I have described a few examples of
typical problems in nuclear astrophysics. Now I discuss how direct
reactions have been used to attempt solving part of these problems.

\section{Direct reactions in/for nuclear astrophysics}

The number of radioactive beam facilities are growing fast around
the world. Some of these facilities use the fragmentation technique,
with secondary beams in the energy range $E_{Lab}\approx100$
MeV/nucleon. Examples are the facilities in GANIL/France, MSU/USA,
RIKEN/Japan and GSI/\-Germany.  In these facilities, direct reactions
have become the main probe of nuclear structure of exotic nuclear
species. In the following, I will present a short discussion of recent
advances in direct reactions for nuclear astrophysics.

\subsection{Transfer reactions}

Transfer reactions are a well established tool to obtain spin, parities,
energy, and spectroscopic factors of states in a nuclear system.
Experimentally, (d,p) reactions are popular due to the simplicity of
the deuteron. Variations of this method have been proposed by several authors.
Examples  are the {Trojan Horse Method} (due to {\it Gerhard Baur}) and the
{Asymptotic Normalization Coefficients} (due to
{\it Akram Mukhamedzhanov} and {\it Natasha Timofeyuk}). They have been discussed by
{\it Bob Tribble} in this conference.
An advantage of using this technique over direct measurements is to avoid
the treatment of the screening problem.

\subsection{Intermediate energy Coulomb excitation}

Before I go on and discuss the Coulomb dissociation method for
nuclear astrophysics, I will discuss a few, and often neglected,
effects in the theory of Coulomb excitation. In low-energy
collisions the theory is very well understood \cite{AW75}. A large
number of small corrections are now well known in the theory and are
necessary in order to analyze experiments on multiple excitation and
reorientation effects. At the other end, the Coulomb excitation of
relativistic heavy ions is characterized by straight-line
trajectories with impact parameter $b$ larger than the sum of the
radii of the two colliding nuclei. A derivation of relativistic
electromagnetic excitation on this basis was performed by {\it Aage
Winther} and {\it Kurt Alder} \cite{WA79}. Later, it was shown that
a quantum theory for relativistic Coulomb excitation leads to
modifications of the semiclassical results~\cite{Ber88}. In
Refs.~\cite{AB89,BN93} the inclusion of relativistic effects in
semiclassical and quantum formulations of Coulomb excitation was
fully clarified.

Recently, the importance of relativistic effects in Coulomb
excitation of a projectile by a target with charge $Z_{2}$, followed
by gamma-decay, in nuclear reactions at intermediate energies was
studied in details. The Coulomb excitation cross section is given by
\begin{equation}
{\frac{d\sigma_{i\rightarrow f}}{d\Omega}}=\left(
\frac{d\sigma}{d\Omega }\right)
_{\mathrm{el}}\frac{16\pi^{2}Z_{2}^{2}e^{2}}{\hbar^{2}}\sum
_{\pi\lambda\mu}{\frac{B(\pi\lambda,I_{i}\rightarrow
I_{f})}{(2\lambda
+1)^{3}}}\mid S(\pi\lambda,\mu)\mid^{2},\label{cross_2}%
\end{equation}
where $B(\pi\lambda,I_{i}\rightarrow I_{f})$ is the reduced
transition probability of the projectile nucleus, $\pi\lambda=E1,\
E2,$ $M1,\ldots$ is the multipolarity of the excitation, and
$\mu=-\lambda,-\lambda+1,\ldots,\lambda$.

The relativistic corrections
to the Rutherford formula for $\left(
d\sigma /d\Omega\right)  _{\mathrm{el}}$ has been investigated
in ref. \cite{AAB90}.
It was shown that
the scattering angle increases by up to 6\% when relativistic
corrections are included in nuclear
collisions at 100 MeV/nucleon. The effect on the elastic scattering
cross section is even more drastic: up to $13\%$ for center-of-mass
scattering angles around 0-4 degrees.

The orbital integrals
$S(\pi\lambda,\mu$) contain the information about relativistic
corrections. Inclusion of absorption effects in
$S(\pi\lambda,\mu$) due to the imaginary part of an optical
nucleus-nucleus potential where worked out in ref. \cite{BN93}.
These orbital integrals depend on the Lorentz factor
$\gamma=(1-v^{2}/c^{2})^{-1/2}$, with $c$ being the speed of light,
on the multipolarity $\pi\lambda\mu$, and on the adiabacity
parameter $\xi (b)=\omega_{fi}b/\gamma v<1$, where
$\omega_{fi}=\left( E_{f}-E_{i}\right) /\hbar$ is the excitation
energy (in units of $\hbar$) and $b$ is the impact parameter.

A recent study in ref. \cite{Ber03} has shown that at 10 MeV/nucleon the
relativistic corrections are important only at the level of 1\%. At
500 MeV/nucleon, the correct treatment of the recoil corrections  is
relevant on the level of 1\%. Thus the non-relativistic treatment of
Coulomb excitation~\cite{AW75} can be safely used for energies below
about 10 MeV/nucleon and the relativistic treatment with a
straight-line trajectory~\cite{WA79} is adequate above about 500
MeV/nucleon. However at energies around 50 to 100 MeV/nucleon,
accelerator energies common to most radioactive beam facilities
(MSU, RIKEN, GSI, GANIL), it is very important to use a correct
treatment of recoil and relativistic effects, both kinematically and
dynamically.
At these energies, the corrections can add up to 50\%.
These effects were also shown in Ref.~\cite{AB89} for the case of
excitation of giant resonances in collisions at intermediate
energies.

A reliable extraction of useful nuclear properties,
like the electromagnetic response (B(E2)-values, $\gamma$-ray
angular distribution, etc.) from Coulomb excitation experiments at
intermediate energies requires a proper treatment of special
relativity \cite{Ber03,BCG03}. The dynamical relativistic effects have often been
neglected in the analysis of experiments elsewhere (see, e.g. \cite{Glas01}).
The effect is highly non-linear, i.e. a 10\% increase in
the velocity might lead to a 50\% increase (or decrease) of certain
physical observables. A general review of the importance of the relativistic
dynamical effects in intermediate energy collisions has been presented in
ref. \cite{Ber05_work,Ber05}.

\subsection{The Coulomb dissociation method}

I refer to the talk by {\it Tohru Motobayshi} for the latest experimental
applications of the Coulomb dissociation method. The idea is quite simple.
The (differential, or angle integrated) Coulomb breakup cross section
for $a+A\longrightarrow b+c+A$ follows from eq. \ref{cross_2}.
It can be rewritten as
\begin{equation}
{d\sigma_{C}^{\pi\lambda
}(\omega)\over d\Omega}=F^{\pi\lambda}(\omega;\theta;\phi)\ .\
\sigma_{\gamma+a\ \rightarrow\ b+c}^{\pi\lambda}(\omega),\label{CDmeth}
\end{equation}
where $\omega$ is the energy transferred from the relative motion to the
breakup, and $\sigma_{\gamma+a\ \rightarrow\ b+c}^{\pi\lambda}(\omega)$ is the photo nuclear cross
section for the multipolarity ${\pi\lambda}$ and photon energy $\omega$. The
function $F^{\pi\lambda}$ depends on $\omega$, the relative motion energy,
nuclear charges and radii, and the scattering angle $\Omega=(\theta,\phi)$.
$F^{\pi\lambda}$
can be reliably calculated \cite{Ber88} for each
multipolarity ${\pi\lambda}$. Time reversal allows one to deduce the radiative
capture cross section $b+c\longrightarrow a+\gamma$ from $\sigma_{\gamma+a\ \rightarrow\ b+c}%
^{\pi\lambda}(\omega)$. This method was proposed by {\it Baur, Bertulani} and
{\it Rebel}, ref. \cite{BBR86}. It has
been tested successfully in a number of reactions of interest for astrophysics.
The most celebrated case is the
reaction $^{7}$Be$($p$,\gamma)^{8}$B, first studied by Motobayashi and
collaborators \cite{Tohru}, followed by numerous
experiments in the last decade. For a recent discussion of the results
obtained with the method, see e.g. ref. \cite{EBS05}.

Motobayashi's experiment using the Coulomb dissociation method immediately
raised the interest of {\it John Bahcall} because the cross section for the
 $^{7}$Be$($p$,\gamma)^{8}$B is a crucial input in John's Standard Solar Model
\cite{Bah89}. In the wake of Motobayshi's
experiment, John sent me four
handwritten letters asking for details of the  method, to which I replied all
by e-mail. Soon the discussion spread among several people. A curious article
entitled "Electronic Battle over Solar Neutrinos", with a partial description
of this discussion, was published in the Science magazine \cite{science}.
John was very happy that new methods had been found to
access information on reactions of astrophysical interest. His quick interest for
the subject was typical of his enthusiasm with new developments in science
and, in particular, in nuclear astrophysics.

Eq. \ref{CDmeth} is based on first-order perturbation theory. It
also assumes that the nuclear contribution to the breakup is small,
or that it can be separated under certain experimental conditions.
The contribution of the nuclear breakup has been examined by several
authors (see, e.g. \cite{BG98}). $^8$B has a small proton separation
energy ($\approx 140$ keV). For such loosely-bound systems it had
been shown that multiple-step, or higher-order effects, are
important \cite{BC92}. These effects occur by means of
continuum-continuum transitions. Detailed studies of dynamic
contributions to the breakup were explored in refs.
\cite{BBK92,BB93} and in several other publications which followed.
The role of higher multipolarities (e.g., E2 contributions
\cite{Ber94,GB95,EB96} in the reaction $^{7}$Be$($p$,\gamma)^{8}$B)
and the coupling to high-lying states has also to be investigated
carefully. In the later case, a recent work has shown that the
influence of giant resonance states is small \cite{Ber02}. It is
worthwhile mentioning that much of the theoretical advances in
understanding the role of the nuclear interaction and of
higher-order effects is due to the work of {\it Stefan Typel, Angela
Bonaccorso, Gerhard Baur, Daniel Baye, Felipe Canto, Radhey Shyam},
and their collaborators.

\subsection{Charge exchange reactions}

During core collapse, temperatures and densities are high enough to
ensure that nuclear statistical equilibrium  is achieved. This
means that for sufficiently low entropies, the matter composition is
dominated by the nuclei with the highest binding energy for a given
$Y_{e}$. Electron capture reduces $Y_{e}$, driving the nuclear
composition to more neutron rich and heavier nuclei, including those
with $N>40$, which dominate the matter composition for densities
larger than a few $10^{10}$~g~cm$^{-3}$. As a consequence of the
model applied in collapse simulations, electron capture on nuclei
ceases at these densities and the capture is entirely due to free
protons. To understand the whole process it is necessary to obtain
Gamow-Teller matrix elements which are not accessible in beta-decay
experiments. Many-body theoretical calculations are right now the
only way to obtain the required matrix elements. This situation can
be remedied experimentally by using charge-exchange reactions.
Charge exchange reactions induced in (p,n) reactions are often used to obtain
values of Gamow-Teller matrix elements, $B(GT)$, which cannot be extracted from
beta-decay experiments. This approach relies on the similarity in spin-isospin
space of charge-exchange reactions and $\beta$-decay operators. As a result of
this similarity, the cross section $\sigma($p,\ n$)$ at small momentum
transfer $q$ is closely proportional to $B(GT)$ for strong transitions
\cite{Tad87}. {\it Taddeucci}'s formula reads
\begin{equation}
{d\sigma\over dq}(q=0)=KN_D|J_{\sigma\tau}|^2 B(\alpha)
,
\end{equation}
where $K$ is a kinematical factor, $N_D$ is a distortion factor (accounting for
initial and final state interactions), $J_{\sigma\tau}$ is the Fourier transform
of the effective nucleon-nucleon interaction, and $B(\alpha=F,GT)$ is the reduced transition
probability for non-spin-flip, $B(F)=
(2J_i+1)^{-1}| \langle f ||\sum_k  \tau_k^{(\pm)} || i \rangle |^2$,
and spin-flip,
$B(GT)=
(2J_i+1)^{-1}| \langle f ||\sum_k \sigma_k \tau_k^{(\pm)} || i \rangle |^2$, transitions.

Taddeucci's formula, valid for one-step processes, was proven to work rather well
for (p,n) reactions (with
a few exceptions). For heavy ion reactions the formula might not work so well.
This has been investigated in refs. \cite{Len89,Ber93,BL97}. In ref. \cite{Len89} it
was shown that multistep processes involving the physical exchange of a proton and
a neutron can still play an important role up to bombarding energies of 100 MeV/nucleon.
Refs. \cite{Ber93,BL97} use the isospin terms of the effective interaction to show that
deviations from the Taddeucci formula are common under many circumstances.
As shown in ref. \cite{Aus94}, for important GT transitions
whose strength are a small fraction of the sum rule the direct relationship
between $\sigma($p,\ n$)$ and $B(GT)$ values also fails to exist. Similar
discrepancies have been observed \cite{Wat85} for reactions on some odd-A
nuclei including $^{13}$C, $^{15}$N, $^{35}$Cl, and $^{39}$K and for
charge-exchange induced by heavy ions \cite{BL97,St96}.
In summary, it is still an open question if Taddeucci's formula is valid in general.

Undoubtedly, charge-exchange reactions such as (p,n), ($^{3}$He,t)
and heavy-ion reactions (A,A$\pm$1) can provide information on the
$B(F)$ and $B(GT)$ values needed for astrophysical purposes. This
will certainly be one of the major research areas in radioactive
beam facilities. This project needs to include an active
collaboration for the reaction-theory part of this method. A very
promising project lead by {\it Remco Zegers} and {\it Sam Austin} at
the NSCL/MSU, using charge-exchange reactions for astrophysical
purposes, is currently under way.

\subsection{Knock-out reactions}

Exotic nuclei are the raw
materials for the synthesis of the heavier
elements in the Universe, and are of considerable
importance in nuclear astrophysics. Modern
shell-model calculations are also now able to
include the effects of residual interactions
between pairs of
nucleons, using forces that reproduce the
measured masses, charge radii and low-lying
excited states of a large number of nuclei.
For very exotic nuclei the small additional
stability that comes with the filling of a particular
orbital can have profound effects upon their
existence as bound systems, their lifetimes and
structures. Thus, verifications of the ordering,
spacing and the occupancy of orbitals are
essential in assessing how exotic nuclei evolve in
the presence of large neutron or proton
imbalance and our ability to predict these
theoretically. Such spectroscopy of the states of
individual nucleons in short-lived nuclei uses
direct nuclear reactions.  The relentless
work of {\it P. Gregers Hansen} on knockout
reactions was one of the most beautiful chapters
of modern nuclear physics. Quoting Gregers \cite{Gregers}:
``Neutron saturated nuclei are the closest one can
get to having a neutron star in the laboratory. The study of
drip-line nuclei has progressed remarkably by observing nuclear
reactions caused by radioactive fragments.''

Single-nucleon knockout reactions with heavy ions, at
intermediate energies and in inverse kinematics, have become a
specific and quantitative tool for studying single-particle
occupancies and correlation effects in the nuclear shell model.
The experiments observe reactions in which fast,
mass $A$, projectiles collide peripherally with a light nuclear
target producing residues with mass $(A-1)$.
The final state of the target and that of the struck nucleon are
not observed, but instead the energy of the final state of the
residue can be identified by measuring coincidences with decay
gamma-rays emitted in flight.

The early interest in knockout reactions came from studies of
nuclear halo states, for which the narrow momentum distributions
of the core fragments
in a qualitative way revealed the large spatial extension of the
halo wave function. It was shown by {\it Bertulani and McVoy}
\cite{ber92} that the longitudinal component of the momentum
(taken along the beam or $z$ direction) gave the most accurate
information on the intrinsic properties of the halo and that it
was insensitive to details of the collision and the size of the
target. In contrast to this, the transverse distributions of the
core are significantly broadened by diffractive effects and by
Coulomb scattering. For experiments that observe the nucleon
produced in elastic breakup, the transverse momentum is entirely
dominated by diffractive effects, as illustrated \cite{ann94} by
the angular distribution of the neutrons from the reaction
$^{9}$Be($^{11}$Be,$^{10}$Be+n)X. In this case, the width of the
transverse momentum distribution reflects essentially the size of
the target.

To
understand the measured longitudinal momentum distributions it is
necessary to take into account that a heavy-ion knockout reaction,
being surface-dominated, can only sample the external part of the
nucleon wave function. The magnitude of the reaction cross section
is determined by the part of the wave function that is accessed,
and the shape of the momentum distribution reflects the momentum
content in this part. Calculations \cite{han95,han96,esb96} based
on a sharp-surface strong-absorption (\textquotedblleft
black-disk\textquotedblright) model could account for the observed
longitudinal momentum distributions and also, approximately, for
the absolute cross sections. This approach is confirmed in the,
more accurate work of {\it Bertulani and Hansen} \cite{BH04}, which
extends the theory to include
the general dependence of the differential cross section on the
momentum vector.
The work of {\it Mahir Hussein} and {\it Kirk McVoy} \cite{HM85}, {\it Angela
Bonaccorso} and {\it David Brink} \cite{BB91},
{\it Kai Hencken, Henning Esbensen} and
{\it George Bertsch} \cite{esb96,hen96} and
of {\it Jim Alkhalili} and {\it Jeff Tostevin} \cite{alk96,tos97,
tos99,tos01,han03}
were crucial for the development of theoretical tools for knock-out
reactions.

\section{Reconciling nuclear structure with nuclear reactions}

Many reactions of interest for nuclear astrophysics involve nuclei
close to the dripline. To describe these reactions, a knowledge of
the structure in the continuum is a crucial feature. Recent work by
{\it Alexander Volya} and {\it Vladimir Zelevinsky} \cite{VZ05},
{\it Nicolas Michel, Witek Nazarewicz, Marek Ploszajczak, Karim
Bennaceur} \cite{Mic04}, and collaborators, are paving the way
toward a microscopic understanding of the many-body continuum. One
basic theoretical problem is to what extent we know the form of the
effective interactions for threshold states. It is also hopeless
that these methods can be accurate in describing high-lying states
in the continuum. In particular, it is not worthwhile to pursue this
approach to describe direct nuclear reactions.

One immediate goal can be achieved in the coming years by using the
Resonating Group Method (RGM) or the
Generator Coordinate Method (GCM). These are a set of coupled integro-differential
equations of the form
\begin{equation}
\sum_{\alpha'} \int d^3 r'
\left[
H^{AB}_{\alpha\alpha'}({\bf r,r'})-EN^{AB}_{\alpha\alpha'}({\bf r,r'})
\right]
g_{\alpha'}({\bf r'})=0,\label{RGM}
\end{equation}
where $H^{AB}_{\alpha\alpha'}({\bf r,r'})=\langle \Psi_A(\alpha,{\bf r})|H|
\Psi_B(\alpha',{\bf r'}) \rangle$ and $N^{AB}_{\alpha\alpha'}({\bf r,r'})
=\langle \Psi_A(\alpha,{\bf r})|
\Psi_B(\alpha',{\bf r'}) \rangle$. In these equations $H$ is the Hamiltonian for the
system of two nuclei (A and B) with the energy $E$, $\Psi_{A,B}$ is the wavefunction
of nucleus A (and B), and $g_{\alpha}({\bf r})$ is a function to be found by numerical
solution of eq. \ref{RGM}, which describes the relative motion of A and B in channel
$\alpha$.
Full antisymmetrization between nucleons of A and B are implicit.
Modern nuclear shell-model calculations, including the No-Core-Shell-Model (NCSM) are able to
provide the wavefunctions $\Psi_{A,B}$ for light nuclei. But the Hamiltonian involves
an effective interaction in the continuum between the clusters A and B. It is
not possible to obtain this effective interaction within the NCSM presently.
Great progress in the microscopic theory of nuclear reactions has been obtained by
{\it Pierre Descouvemont} and
{\it Daniel Baye}. More applications of the RGM method using NCSM wavefunctions
was presented by {\it Christian Forssen} in this conference.

Overlap integrals of the type $I_{Aa}(r)=\langle
\Psi_{A-a}|\Psi_A\rangle$ for bound states has been calculated by
{\it Petr Navratil} \cite{Petr04} within the NCSM. This is one of
the inputs necessary to calculate S-factors for radiative capture,
$S_\alpha \sim |\langle g_{\alpha}|H_{EM}|I_{Aa}\rangle|^2$, where
$H_{EM}$ is a corresponding electromagnetic operator. The left-hand
side of this equation is to be obtained by solving eq. \ref{RGM}.
For some cases, in particular for the p$+^7$Be reaction, the
distortion caused by the microscopic structure of the cluster does
not seem to be crucial to obtain the wavefunction in the continuum.
The wavefunction is often obtained by means of a potential model.
The NCSM overlap integrals, $I_{Aa}$, can also be corrected to
reproduce the right asymptotics \cite{NBC05}, given by
$I_{Aa}(r)\propto W_{-\eta,l+1/2}(2k_0r)$, where $\eta$ is the
Sommerfeld parameter, $l$ the angular momentum, $k_0=\sqrt{2\mu
E_0}/\hbar$ with $\mu$ the reduced mass and $E_0$ the separation
energy.

A step in the direction of reconciling structure and reactions for
the practical purpose of obtaining astrophysical S-factors, along
the lines described in the previous paragraph, was obtained in ref.
\cite{NBC05,NBC06}. The wavefunctions obtained in this way were
shown to reproduce very well the momentum distributions in knockout
reactions of the type $^8$B$+A\longrightarrow \ ^7$Be$+X$ obtained
in experiments at MSU and GSI facilities. The astrophysical S-factor
for the reaction  $^{7}$Be$($p$,\gamma)^{8}$B was also calculated
and excellent agreement was found with the experimental data in both
direct and indirect measurements \cite{NBC05,NBC06}. The low- and
high-energy slopes of the  S-factor obtained with the NCSM is well
described by the fit
\begin{equation}
S_{17}(E)=(22.109\ {\rm eV.b}){1+5.30E+1.65E^2+0.857E^3 \over 1+E/0.1375}  ,
\end{equation}
where E is the relative energy (in MeV) of p$+^7$Be in their
center-of-mass. This equation corresponds to a Pad\'e approximant of
the S-factor. A subthreshold pole due to the binding energy of $^8$B
is responsible for the denominator \cite{JKS98,WK81}.
\section{Future}

Extremely exciting experimental results on direct reactions in/for
nuclear astrophysics will be produced in the future. New radioactive
beam facilities will be constructed around the world. Among the
several proposed experiments, I mention the R3B (contact person:
{\it Thomas Aumann}) and the ELISE (contact person: {\it Haik
Simon}) projects, both at the future FAIR facility in GSI. The first
project will use radioactive beams and direct reactions to obtain
the nuclear physics input for astrophysics. The ELISE experimental
setup will use electrons scattered off radioactive nuclei. These
experiments will explore an unknown world of studies with nuclei far
from stability which play an important role in our universe.

The US needs urgently a new radioactive beam facility, fully
dedicated to the physics of radioactive nuclei. Without competing
facilities worldwide, observational and theoretical astrophysics
will never be able to constrain numerous models used to understand
our universe. And without new inputs and constraints set
by nuclear physics, astrophysics would slowly become a "no-man's
land" (or "all-you-can-eat") science.

\end{document}